\documentclass[conference]{IEEEtran}
\IEEEoverridecommandlockouts
\usepackage{epstopdf}
\usepackage[draft]{hyperref} 
\usepackage{setspace}
\usepackage{amsmath}
\usepackage{amssymb}
\usepackage{amsthm}
\usepackage{amsfonts}
\usepackage{color}
\usepackage{graphicx}
\usepackage{subfigure}  
\usepackage{cite}
\usepackage{mathtools}
\usepackage{stmaryrd}
\usepackage[mathscr]{euscript}
\usepackage{mathrsfs}

\usepackage{soul}

\usepackage{algorithm}
\usepackage{algpseudocode}
\usepackage{float}

\begin{document}
\title{Super-Resolution via Learned Predictor 
\thanks{$^*$These authors contributed equally to this work.\\The authors would like to thank Dr. Sridevi Gutta, (Late) Prof. H. S. Atrey, and Dr. Prasad Sudhakar for their discussions on formulating the work.}
}
%
\author{\IEEEauthorblockN{Sampath Kumar Dondapati $^{1,}$$^{2,}$$^*$, Omkar~Nitsure$^{2,}$$^*$, Satish~Mulleti$^{2}$}
\IEEEauthorblockA{$^{1}$\textit{Satish Dhawan Space Center-SHAR, ISRO}, India. \\$^{2}$ \textit{Department of Electrical Engineering, Indian Institute of Technology (IIT) Bombay}, India.\\ 
Emails: mr.sampathkumar.d@gmail.com, omkarnitsure2003@gmail.com, 
mulleti.satish@gmail.com}}


%
%
%
\maketitle
\begin{abstract}
Frequency estimation from measurements corrupted by noise is a fundamental challenge across numerous engineering and scientific fields. Among the pivotal factors shaping the resolution capacity of any frequency estimation technique are noise levels and measurement count. Often constrained by practical limitations, the number of measurements tends to be limited. This work introduces a learning-driven approach focused on predicting forthcoming measurements based on available samples. Subsequently, we demonstrate that we can attain high-resolution frequency estimates by combining provided and predicted measurements. In particular, our findings indicate that using just one-third of the total measurements, the method achieves a performance akin to that obtained with the complete set. Unlike existing learning-based frequency estimators, our approach's output retains full interpretability. This work holds promise for developing energy-efficient systems with reduced sampling requirements, which will benefit various applications. 
\end{abstract}
\section{Introduction}
\label{sec:intro}
 The task of frequency estimation is to determine a set of $L$ frequencies from a set of $M$ measurements. This is a common problem in various applications, such as direction-of-arrival estimation, Doppler estimation in radar and sonar, ultrasound imaging, optical coherence tomography, and nuclear magnetic spectroscopy. In most applications, the frequency resolution of algorithms is a crucial factor and generally increases with the number of measurements in the presence of noise. However, the number of measurements is often limited due to cost and power consumption. Hence, algorithms capable of achieving high resolution with fewer measurements are desirable.


Theoretically, $2L$ samples are sufficient and necessary to determine $L$ frequencies uniquely in the absence of noise. In practice, Prony's algorithm can operate with such limited samples \cite{prony}. Though the algorithm has infinite resolution in the absence of noise, it is highly unstable at the slightest noise level. Denoising algorithms could be applied to improve the noise robustness at the expense of high computations \cite{cadzow,condat_cadzow}.

The investigation centered on high-resolution spectral estimation (HRSE) has led to the exploration of multiple noise-resistant techniques (detailed in Chapter 4 of \cite{stoica}). High resolution in this context denotes algorithmic precision that surpasses that of the periodogram or Fourier transform-based approach, characterized by $1/M$. Among these techniques is the multiple signal classification method (MUSIC) \cite{barabell_rmusic, schmidt_music, bdrao_rmusic}, the estimation of signal parameters using the rotational invariance technique (ESPRIT) \cite{paulraj_esprit}, and the use of the matrix pencil method \cite{sarkar_mp}. These techniques exhibit superior robustness and enhanced resolution capabilities compared to periodogram and annihilating filter methods. In \cite{bdrao_rmusic}, a precise mathematical expression is derived for frequency estimation, highlighting the inverse relationship between error and minimum frequency separation under medium to high signal-to-noise ratios (SNRs) as $M$ tends to infinity. High resolution might necessitate a sufficiently large value of $M$ in scenarios with non-asymptotic conditions and low SNR.

The majority of the aforementioned HRSE techniques covered in the previous section are built upon uniform consecutive samples of sinusoidal signals. Multiple strategies have emerged for estimating frequencies from irregular samples \cite{stoica_nu_2009, babu2010spectral, tang2013compressed, bhaskar2013atomic, yang2015gridless}. In particular, \cite{tang2013compressed, bhaskar2013atomic, yang2015gridless} introduces an approach centered on minimization of atomic norms, which extends the concept of a norm $\ell_1$ for the estimation of sparse vectors \cite{candes_spmag, eldar_cs_book}. Instead of using all $M$ consecutive samples of the signal, the atomic norm-based method estimates frequencies with high probability from $\mathcal{O}(L, \log L, \log M)$ random measurements from available $M$ measurements, provided that no two frequencies are closer than $4/M$ \cite{tang2013compressed, bhaskar2013atomic}. Though the theoretical resolution ability of the method is four times lower than the periodogram, in practice, the algorithm is shown to resolve frequencies separated by $1/M$. However, these techniques have high computational complexity due to reliance on semidefinite programming.

A potential reduction in computational burden is attainable through data-driven methodologies. For instance, \cite{izacard2019learning} introduced PSnet, a deep-learning-based scheme for frequency estimation. Analogous to the principles behind periodogram and root-MUSIC \cite{bdrao_rmusic}, the authors suggest an initial step of learning a pseudo-spectrum from samples, followed by frequency estimation via spectrum peak identification. Empirical observations reveal the superiority of PSnet over MUSIC and periodogram when the minimum frequency separation equals $1/M$. The paper's important insight is that estimating frequencies from a learned frequency representation (e.g., pseudo-spectrum) is advantageous instead of training a deep network to directly learn frequencies from the samples, as attempted in \cite{guo_doa_2020}. An enhanced version of PSnet is proposed in \cite{data_driven_learning}, introducing architectural improvements in the network and incorporating a learned estimator for the frequency count $L$. These techniques employ triangular and Gaussian kernels to construct pseudo-spectra, although the optimal kernel selection remains an open question. Furthermore, the impact of pseudo-spectrum discretization on learning was not addressed. Diverse adaptations and extensions of these approaches are presented in \cite{wu_2019, elbir2020deepmusic, xie_damped_2021, pan_learn_2022, chen2022sdoanet, ali_doa_learning_2023,smith2024frequency,biswas2024deep}. In all these approaches, the choice of pseudo-spectrum or spectral representation, which is learned by the network, is contentious both in terms of its choice during training and its interpretability. Notably, the approaches were trained and tested with examples where the minimum frequency separation was $1/M$. The performance of these approaches for resolutions of less than $1/M$ is not evaluated.

Several alternative learning strategies assume a grid-based arrangement of frequencies and employ multilabel classification to tackle the issue \cite{papageorgiou_deep_doa_2021}. This ``on-grid'' presumption, however, imposes limitations on resolution capabilities. In a different direction, a subset of methods learns denoising techniques before applying conventional frequency estimation methods \cite{jiang_denoise_2019, leung_fri_learn_2023}. These approaches necessitate broad training across various noise levels \cite{jiang_denoise_2019} or demand an extensive set of training samples \cite{leung_fri_learn_2023}. In a nutshell, a non-learning-based approach requires many samples to reach a desired resolution in the presence of noise. However, the learning-based methods are data-hungry, and non-interpretable, and the resolution is still not below the periodogram limit.

In this paper, we adopt a novel data-driven approach for frequency estimation from noisy measurements to achieve high resolution while keeping the interpretability. Any frequency estimation algorithm's behavior reveals that, for a fixed resolution, estimation errors decrease with a higher signal-to-noise ratio (SNR) and/or an increased number of samples. Denoising can mitigate noise effects, although it possesses inherent limitations \cite{cadzow, leung_fri_learn_2023}. Consequently, our focus shifts towards enhancing the number of samples, aiming to boost accuracy. Our proposed solution entails a learning-based predictor that augments the number of samples by extrapolating from a small set of noisy samples. Specifically, we predict $N-M$ samples from $M$ measurements, where $N>M$. By leveraging the provided $M$ samples alongside the predicted ones, we approach the accuracy achievable with $N$ samples. We propose two architectures for the learnable predictor. The first one is based on a long-short-term-memory (LSTM) convolution neural network (CNN), whereas the second one is realized by a transformer (TF) encoder \cite{vaswani2017attention}. Unlike methods such as \cite{izacard2019learning, data_driven_learning, wu_2019, elbir2020deepmusic, xie_damped_2021, pan_learn_2022, chen2022sdoanet, ali_doa_learning_2023}, our approach does not require the selection or design of pseudo-spectra which may not have an interpretation in terms of conventional Fourier spectra. Through simulations, we show that by using $M$ noisy measurements, we achieve a higher resolution and lower errors for different SNRs than the HRSE approach and the method in \cite{data_driven_learning}, especially with the TF-based predictor.

The structure of the paper unfolds as follows: Section 2 defines the signal model and outlines the problem formulation. In Section 3, we delve into the details of the proposed predictor. Section 4 covers network design and simulation results, followed by conclusions.

\section{Problem Formulation}
Consider $N$ uniform samples of a linear combination of complex exponentials given as
\begin{align}
x(n) = \sum_{\ell=1}^L a_{\ell} \, \exp(\mathrm{j} 2\pi f_\ell \, n)+ w(n), \quad n =1, \cdots, N. \label{eq:fn}
\end{align}
Here, the coefficients $a_\ell \in \mathbb{R}$ represent amplitudes, and $f_\ell \in (0, 0.5]$ represent normalized frequencies. The term $w(n)$ is additive noise, which is assumed to be a zero-mean Gaussian random variable with variance $\sigma^2$, with the samples being independent and identically distributed. The objective is to estimate the frequencies $\{f_\ell\}_{\ell=1}^L$ from the measurements.

Estimation accuracy relies on the following factors: the SNR, measured in decibels as $10\log_{10} \left(\frac{|x(n)|_2^2}{N\sigma^2}\right)$, number of measurements $M$, and resolution $\Delta_f$, representing the smallest distance between any two frequencies in $x(n)$. Keeping SNR and resolution constant means that higher accuracy requires a larger number of measurements. On the other hand, for an acceptable estimation accuracy and a given SNR, the achievable resolution is around $1/N$ for HRSE methods if all the $N$ measurements are used. When it is expensive to gather a large number of measurements and a subset of the measurements $\{x(n)\}_{n=1}^M$ are available, where $M<N$, the resolution decreases to $1/M$. 

The question is whether a resolution of $1/N$ lower is achievable from $M$ measurements. We show that this is possible by using a learning-based approach for frequency estimation, which will be discussed in the following sections.

\section{Learnable Predictor}
Utilizing HRSE and non-learnable techniques to estimate frequencies from the complete set of $N$ measurements might yield favorable accuracy and resolution. Nonetheless, this may not hold true in cases where only $M$ (consecutive) samples are accessible. We introduce a two-step strategy to achieve resolution closer to $1/N$ using the available $M$ measurements. In the first stage, we predict samples $\{x(n)\}_{n = M+1}^N$ based on the provided measurements $\{x(n)\}_{n=1}^M$. The predicted samples are represented as $\{\Tilde{x}(n)\}_{n = M+1}^N$. Subsequently, we concatenate these estimated samples with the existing ones and employ HRSE methods to estimate the frequencies.

\begin{figure}[!t]
    \centering
    \includegraphics[width=2.5in]{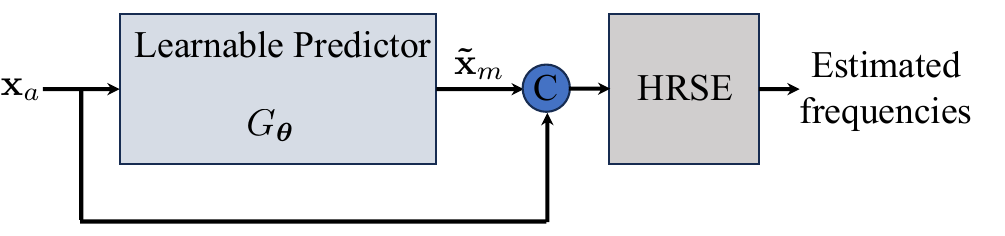}
    \caption{Schematic of the proposed frequency estimation method: The predictor estimates the future samples from the given samples and then concatenates with the true samples by using concatenation operation C. Then HRSE methods are used to estimate frequencies from the samples.}
    \label{fig:block}
\end{figure}

In the initial phase, when noise is absent, it is possible to achieve perfect prediction for $x(n)$ \cite{vaidyanathan2022lp}. Specifically, in the absence of noise, for any $x(n)$ as in \eqref{eq:fn}, there exists a set of coefficients $\{c_k\}_{k=1}^K$ such that $\displaystyle x(n) = \sum_{k=1}^{K} c_k \, x(n-k), n = K+1, \cdots, N$, where $K \geq L$. In the presence of noise, $\{c_k\}_{k=1}^L$ are estimated using least-squares, aiming to minimize the error $ \sum_{n=L+1}^N \left|x(n) - \sum_{k=1}^L c_k , x(n-k) \right|^2$. However, this predictive approach remains susceptible to errors in low SNR and situations with lower resolution. Furthermore, the coefficients have to be recalculated for different signals.

To construct a robust predictor applicable to a class of signals, we train deep learning models to estimate $\{{x}(n)\}_{n = M+1}^N$ based on the available samples. To elaborate, consider the vectors $\mathbf{x}_a \in \mathbb{R}^M$ and $\mathbf{x}_m \in \mathbb{R}^{N-M}$, representing the available first $M$ measurements and the last $N-M$ samples, respectively. We design a predictor $G_{\boldsymbol{\theta}}: \mathbb{R}^M \rightarrow \mathbb{R}^{N-M}$, parameterized by $\boldsymbol{\theta}$, with the objective of minimizing the mean-squared estimation error $\|\mathbf{x}_m - G_{\boldsymbol{\theta}}(\mathbf{x}_a)\|_2^2$ across a set of examples. To be precise, let $\{\mathbf{x}_i\}_{i=1}^I$ represent a collection of noisy samples following the form in \eqref{eq:fn}, where frequencies and amplitudes vary among examples. By splitting the measurements of these examples into the initial $M$ measurements and the remaining set, we obtain the vectors $\{\mathbf{x}_{a,i}\}_{i=1}^I$ and $\{\mathbf{x}_{m,i}\}_{i=1}^I$. The network parameters $\boldsymbol{\theta}$ are optimized to minimize the cost function
\begin{align}
\sum_{i=1}^I \|\mathbf{x}_{m,i} - G_{\boldsymbol{\theta}}(\mathbf{x}_{a,i})\|_2^2.
\end{align}

During the inference, for any set of $M$ noisy samples $\mathbf{x}_a$, the future samples are predicted as $\mathbf{\tilde{x}}_{m} = G_{\boldsymbol{\theta}}(\mathbf{x}_a)$. Then a HRSE method is applied on the concatenated vector $[\mathbf{x}_a^{\mathrm{T}}\,\, \mathbf{\tilde{x}}_{m}^{\mathrm{T}}] ^{\mathrm{T}}$ for frequency estimation. The overall system is shown in Fig.~\ref{fig:block}.




\section{Experimental Design and Simulation Results}
In this section, we first discuss network design and data generation. Then, we compare the proposed approach with the existing methods.
\subsection{Network Architecture}
 We first discuss the CNN-based architecture and then discuss the TF model.
 
The CNN-based architecture of the learnable predictor is a hybrid deep learning architecture that combines convolutional and recurrent neural network layers. It begins with three stacked Conv1D layers, each with 64 filters and kernel sizes of 5, 7, and 9, respectively, designed to capture temporal patterns in the input signal. Batch normalization and dropout layers are applied after each convolutional layer to enhance model generalization and prevent overfitting. Following the CNN layers, a Bidirectional LSTM layer with 64 units captures sequential dependencies in both forward and backward directions, further improving temporal pattern recognition. The model concludes with a fully connected dense layer with linear activation to produce the final output.


The TF-Encoder-based architecture consists of multi-layer perceptrons for projecting the signal to a higher dimensional space (transformer feature dimension of 64), followed by batch normalization for training stabilization. We employed a positional encoding layer to incorporate the relative positional significance. The transformer encoder was used as the backbone of the architecture with a feedforward network dimension of 1024 and two layers. A linear layer (expansion layer) was used to predict $N-M$ samples from $M$ samples, and another linear layer was used to project back to the signal space. MLPs in Transformer use ReLU activation while all other MLPs use a Leaky ReLU activation with a slope of 0.2 \cite{vaswani2017attention}. The model we use has 0.46 million learnable parameters. 

\subsection{Training Data Generation}
\label{sec:training_data}
Since the task is to improve the resolution from $1/M$ to $1/N$, we considered only two frequencies ($L=2$) with equal amplitudes during training and testing. Further, to have a resolution of at least $ \Delta_f = 1/N$ and potentially improve on smaller resolutions, the samples were generated such that they had some examples with resolutions smaller than $1/N$. We generated the following six data sets to provide a sufficient mix of frequencies for the model to generalize well.
\begin{itemize}
    \item Set-1: 20000 examples were generated with minimum resolution. For each example, $f_1$ was chosen uniformly at random from the interval $[0, 0.5-\Delta_f]$. Then the second frequency was set as $f_2 = f_1 + 0.5\Delta_f$.  
    \item Set-2: In this set, 20000 examples were generated with a resolution greater than $\Delta_f$. For each example, $f_1$ was chosen randomly as in the previous set. Then, we selected $f_2 = f_1+\Delta_f+\epsilon$ where $\epsilon$ was generated from a uniform distribution $[-(f_1+\Delta_f), (0.5-f_1-\Delta_f)]$. The distribution of $\epsilon$ ensures that $|f_2-f_1|\geq \Delta_f$.
    
    \item Set-3: This set consists of 5625 examples where both frequencies were randomly generated from a two-dimensional grid with a grid size $\Delta_f$. Since the frequencies had to be limited to the range $[0, 0.5]$, for each frequency, the grid points were given by the set $\{k\Delta_f \}_{k=1}^{75}$. 
    \item Set-4: In this set, $f_1$ was chosen as in Set-1 or Set-2, and then, we set the second frequency as $f_2 = f_1+k\Delta_f$ where $k$ was chosen randomly from the integer set $\left[\lceil (-\frac{f_1}{\Delta_f}) \rceil, \cdots, \lfloor \frac{0.5-f_1}{\Delta_f} \rfloor \right]$. A total of 20000 examples were generated in this set.
     \item Set-5: This set consists of 20000 examples in which both $f_1$ and $f_2$ were chosen uniformly at random from the interval $[0, 0.5]$
    \item Set-6: In this set, $f_1$ was chosen from a Gaussian distribution with a mean and standard deviation of 0.25 and restricted to the range $[0, 0.5]$. $f_2$ was chosen uniformly at random from the interval $[0, 0.5]$. A total of 20000 examples were
    generated in this set.
\end{itemize}

Set 3 and 4 were generated to help the network to separate frequencies separated by integer multiples of $\Delta_f$, which is the key in the resolution analysis.
 The above six sets have 105625 possible pairs of frequencies $(f_1, f_2)$. The samples were generated for each pair of frequencies following \eqref{eq:fn} with $a_1= a_2=1$ and for $N=150$. Each of these 105625 signal samples was corrupted with noise for a given SNR. Specifically, three independent noise instances for each example were realized, resulting in a total of 316875 examples. We also trained different networks for each SNR.

By keeping $M=50$ (note that $N=150$ in our setup), the network was trained for 50 epochs with a batch size of 128 in the CNN-based model and 2048 in the TF-based model. The training loss was minimized using the Adam optimizer with a learning rate of 0.001. A performance comparison of the proposed algorithm through simulations will be discussed next.

\subsection{Simulation Results}	
\label{sec:simulations}
In this section, we compared the following methods for frequency estimation by setting $N=150$ and $M=50$.
\begin{itemize}
    \item  {\it 50T}: Used $M = 50$ true noisy samples with ESPRIT \cite{paulraj_esprit}. 
     \item  {\it 150T}: Used $N = 150$ true noisy samples with ESPRIT.
     \item  {\it CNN-50T+100P}: Used 50 true noisy ({\it T}) with 100 predicted ({\it P}) samples with ESPRIT. The predictor is realized by CNN.
   
     \item  {\it TF-50T+100P}: Same as the previous approach, but the prediction is via the TF.
     \item {\it Deep-Freq} model\cite{data_driven_learning}. The model is trained using the proposed dataset
   \end{itemize}



 For an objective comparison, we used normalized mean-squared error (NMSE), which was computed as
 \begin{align}
     \text{NMSE} = \frac{{\frac{1}{K}}\sum_{k=1}^K(f_k - \tilde f_k)^2}{{\frac{1}{K}}\sum_{k=1}^K(f_k)^2},
     \label{eq:mse}
 \end{align}
 where $\tilde f_k$ is an estimate of the frequency $f_k$. 
 
\begin{figure}[!t]
    \centering
    \subfigure[SNR = 5dB]{
    \includegraphics[width= 1.6 in]{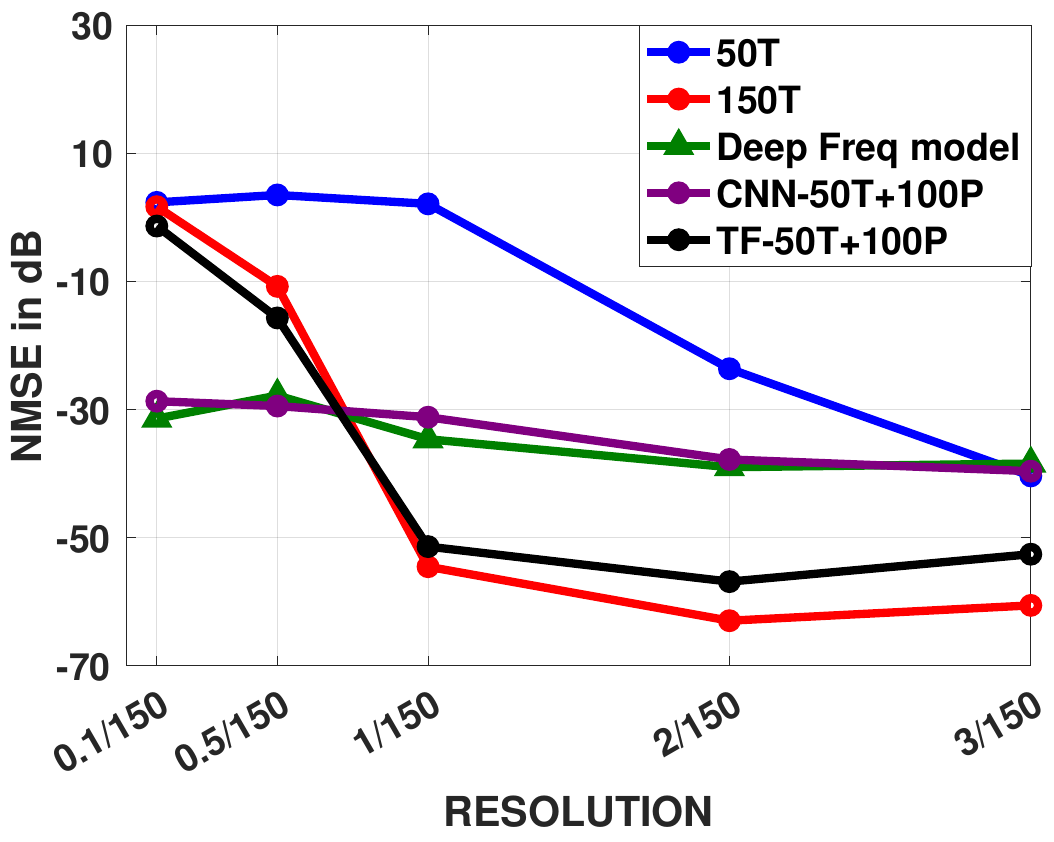}}
    \subfigure[SNR=15 dB]{\includegraphics[width= 1.6 in]{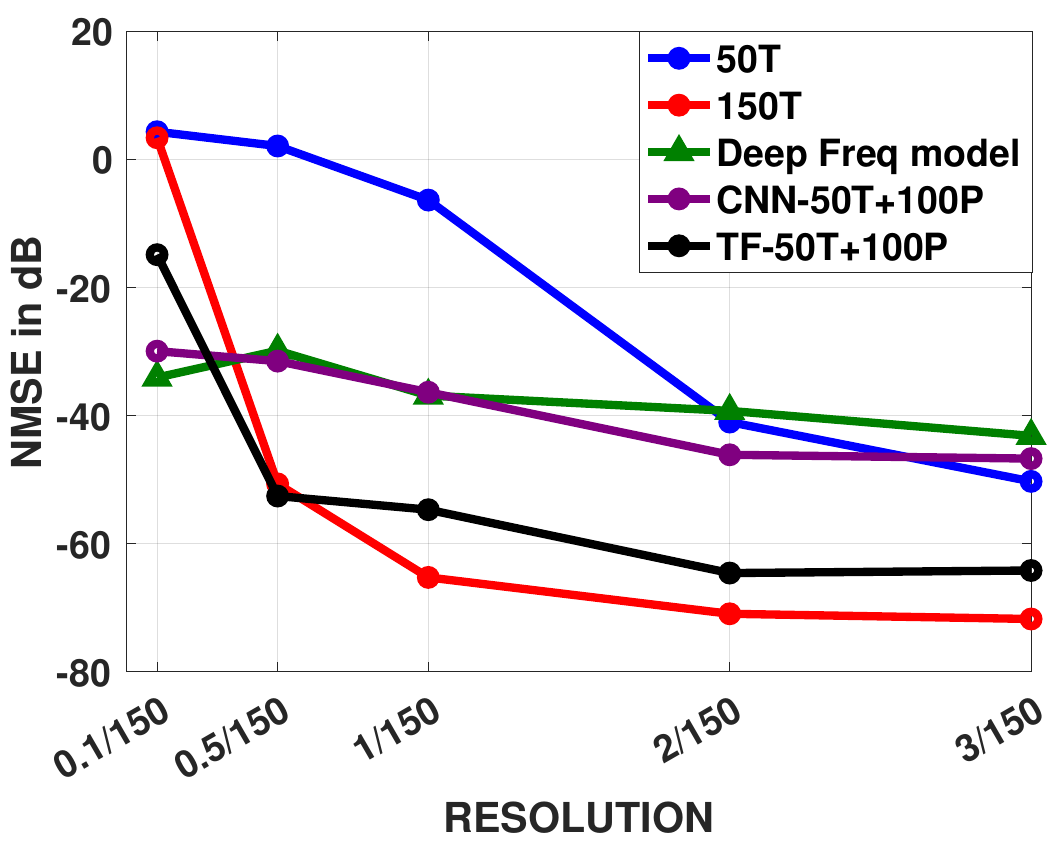}}
    \subfigure[$\Delta_f = 1/150$]{\includegraphics[width= 1.6 in]{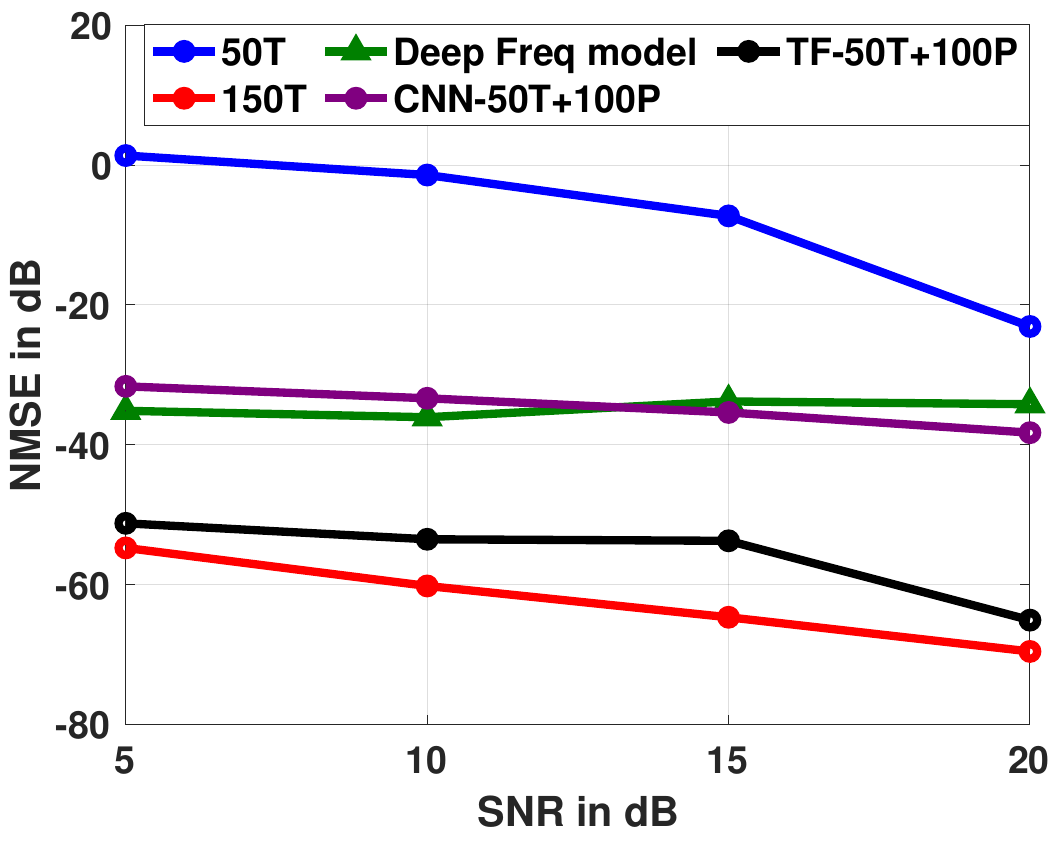}}
    \caption{A comparison of various methods for different resolutions at 5dB and 15dB SNR where $L=2$: The proposed approach, labeled as TF-50T+100P, has $50$ dB lower error than 50T for frequency separation equal to the resolution limit $\Delta_f = 1/150$.}
    \label{fig:resolution_result}
\end{figure}

In the following simulation, we examine the methods under varying resolution conditions for SNRs 5 dB and 15 dB. For each SNR and a given resolution, 200 test examples are generated. For each example, first, $f_1$ is randomly selected and then we set $f_2 = f_1 + \Delta$, where $\Delta$ represents the desired resolution.
With a limited sample size of $M=50$, the Fourier-based periodogram method achieved a resolution threshold of $1/50 = 0.02$ Hz, whereas the full $N$-sample scenario yielded a resolution of $\Delta_f =1/N = 0.0067$ Hz. The key question was whether we could achieve the $1/N$ resolution threshold using only $M$ samples.

In Figs.~\ref{fig:resolution_result}(a) and (b), MSEs for various resolutions are shown for SNRs 5 dB and 15 dB, respectively. Starting from 50 samples, the learning-based approaches were able to have better resolution abilities than {\it 50T}. Among the learnable methods, {\it TF-50T+100P} was able to achieve the same (super)resolution ability as that of {\it 150T}. Interestingly, at SNR 5 dB, the {\it CNN-50T+100P} and {\it DeepFreq} have lower errors compared to {\it 150T} and {\it TF-50T+100P} for resolutions below $\Delta f = 1/150$. But for higher resolutions, the NMSEs for these methods are not decreasing as rapidly as {\it TF-50T+100P} and {\it 150T}.


Next, we compared the methods for different SNRs while keeping the resolution to $\Delta_f = 1/150$, and the NMSEs are depicted in Figs.~\ref{fig:resolution_result}(c). For each SNR, 1000 test samples were used where the frequencies $f_1$ were picked randomly, and then we set $f_2 = f_1 + \Delta_f$. As expected, {\it 50T} results in the highest error while {\it 150T} has around 50 dB less error at the expense of thrice the measurements. In comparison, learning-based methods have significantly lower error than {\it 50T} while using the same number of measurements. Specifically, $15-35$ dB lower for {\it CNN-50T+100P} and {\it DeepFreq}, and $45-50$ dB for {\it TF-50T+100P}.

In a nutshell, the learning-based methods, especially, the proposed {\it TF-50T+100P} is able to achieve super-resolution with far fewer measurements for various SNRs.

\section{Conclusions}	
In this study, we present a novel learning-centered technique for accomplishing super-resolution frequency estimation. Our approach involves training a predictor capable of forecasting upcoming samples in a sum of complex exponential using a limited sample set. By subsequently incorporating both the available and predicted samples, we successfully demonstrate the attainment of high-resolution estimation. This algorithm proves valuable in scenarios where measurements come at a premium. Our ongoing efforts are directed towards expanding the method to predict samples from non-uniform datasets.

\bibliographystyle{IEEEtran}
\bibliography{refs,refs2,refs_learn,refs_learn2}

\end{document}